\documentclass[
    ,final            
  ]
  {aipproc}

\layoutstyle{6x9}

\def\chandra    {{\em Chandra}\/}

\begin{document}

\title{Why are there strong radio AGNs in the center of \\
``non-cool core'' clusters?}


\author{Ming Sun}{
  address={Department of Astronomy, University of Virginia, P.O. Box 400325, Charlottesville, VA 22904-4325; msun@virginia.edu}
}

\begin{abstract}
Radio AGN feedback in X-ray cool cores has been proposed as a crucial ingredient
in the evolution of baryonic structures. However, it has long been
known that strong radio AGNs also exist in ``noncool core'' clusters,
which brings up the question whether an X-ray cool core is always required
for radio feedback. We present a systematic analysis of 152 groups and clusters
to show that every BCG with a strong radio AGN has an X-ray cool core.
Those strong radio AGNs in the center of the ``noncool core'' systems
identified before are in fact associated with small X-ray cool cores with typical
radii of $<$ 5 kpc (we call them coronae). Small coronae are most likely of ISM
origin and they carry enough fuel to power radio AGNs. Our results suggest that the traditional
cool core / noncool core dichotomy is too simple. A better alternative is the
cool core distribution function with the enclosed X-ray luminosity.
Other implications of our results are also discussed, including a warning on
the simple extrapolation of the density profile to derive Bondi accretion rate.
\end{abstract}

\maketitle


\section{Radio AGNs and X-ray cool cores}

The importance of AGN outflows for cosmic structure formation and evolution
has been widely appreciated recently. AGN outflows may simultaneously explain
the antihierarchical quenching of star formation in massive galaxies, the
exponential cut-off at the bright end of the galaxy luminosity function,
the $M_{\rm SMBH} - M_{\rm bulge}$ relation and the quenching of cooling-flows
in cluster cores (e.g., \citep{cro06}; \citep{mn07}), as also widely discussed in this conference.
It has been suggested that FRI and low-excitation FRII radio galaxies are
fueled through accretion of hot gas in the so-called ``radio mode''
\citep{chu05}\citep{best05}\citep{cro06}\citep{allen06}\citep{hard07}.
This naturally brings out the following observational questions: 
{\em What fuels the strong radio AGN in
groups and clusters, especially those not in large cool cores? Is an X-ray cool
core always required for strong radio AGN in groups and clusters?}

\begin{figure}
  \includegraphics[height=.21\textheight]{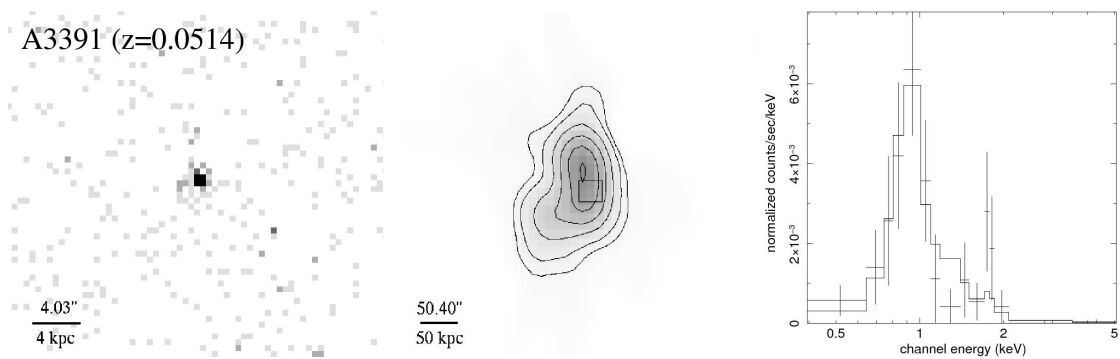}
  \caption{One example of strong radio AGNs in the center of a cluster
without a large X-ray cool core (A3391). \chandra\ image around the BCG is
shown on the left, while the middle panel shows the radio AGN (the small box shows
the field of the left panel). The radio AGN is 2.5 times more luminous than
Perseus's at 1.4 GHz, but the ICM cooling time at $r >$ 4 kpc is 17 Gyr.
The X-ray point-like source centered on the BCG is in fact a small cool core,
revealed by its spectrum shown on the right (clearly a thermal origin).
As shown by \citep{sun09}, the small cool core of A3391 has enough
gas to fuel its radio AGN, although its total luminosity is only 0.062\% of Perseus's
central cool core.
}
\end{figure}

First of all, it has long been known that there are many strong radio AGNs in the center of
``noncool core'' clusters, or clusters without a large cool core.
Some examples are in the HIFLUGCS sample\citep{Mit09}, e.g.,
A3391, A3395s, A2634 and A400. \citep{Mit09} cited cluster merger and other
mechanisms to explain these ``outliers''. 
One example is shown in Fig. 1 (A3391). There is an X-ray point-like source
at the position of the BCG. Is it an X-ray AGN? A spectral analysis clearly reveals
a significant iron-L hump, implying a thermal origin. Its properties are very similar
to many other coronae (or mini-cooling cores) discussed in \citep{sun07} (also see the
references there).
Thus, we have a case in sharp contrast to the Perseus cluster, a much much smaller
X-ray cool core associated with a stronger radio AGN.
Is this case unique? Not at all! \citep{sun09} presents many more similar cases from
a sample study of 152 nearby groups and clusters from the \chandra\ archive.
As shown in the left panel of Fig. 2, all 69 BCGs with a strong radio AGN
($L_{\rm 1.4 GHz} > 2 \times 10^{23}$ W Hz$^{-1}$, no high-excitation FRII galaxies)
have X-ray cool cores with a central isochoric cooling time of $<$ 1 Gyr \citep{sun09}.
This conclusion also holds in the B55 and the extended HIFLUGCS
samples \citep{sun09}. The BCG cool cores can be divided
into two classes, large ($r_{\rm 4 Gyr} >$ 30 kpc) and luminous
($L_{\rm 0.5-2 keV} \geq 10^{42}$ ergs s$^{-1}$) cool cores like Perseus's
cool core, or small ($\leq$ 4 kpc in radius typically) coronae like A3391's.
We call them the large-cool-core (LCC) class (the focus of this conference) and the corona class. The gas of the
former class is primary of ICM origin, while the latter one is of ISM origin.
More detail on these two classes can be found in \citep{sun09}.

\begin{figure}
  \includegraphics[height=.38\textheight]{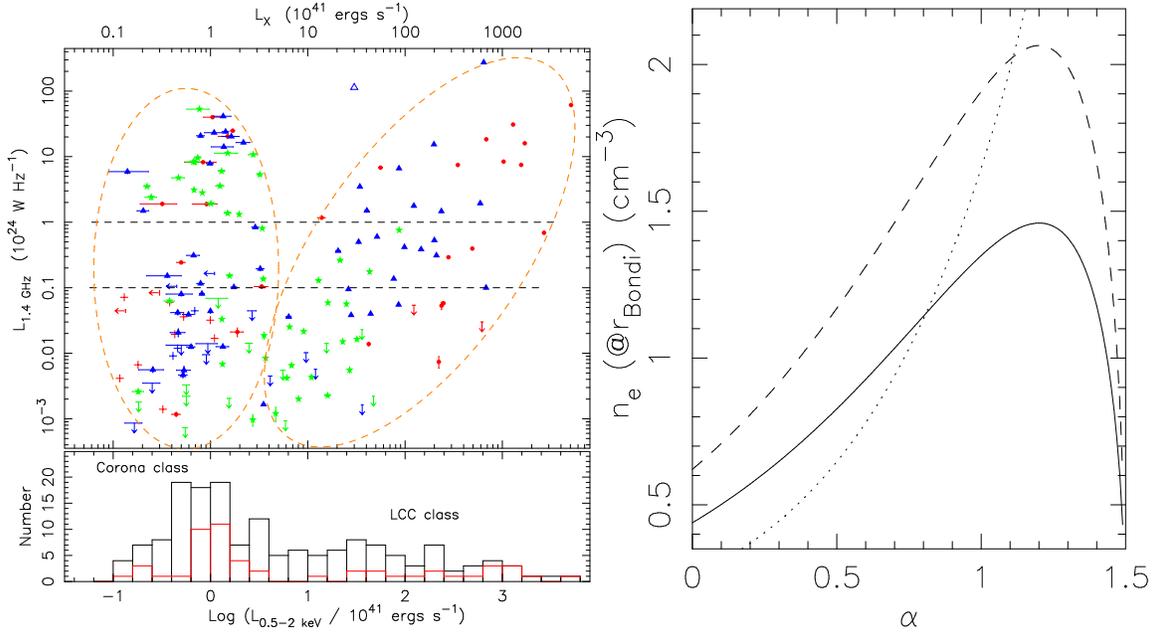}
  \caption{{\bf Left}: The upper panel shows the rest-frame 0.5-2 keV luminosity of the cool core
(within a radius where the cooling time is 4 Gyr) of the BCG vs. the 1.4 GHz
luminosity of the BCG. Red filled circles are for $kT>$4 keV clusters. Blue triangles are
for $kT$=2-4 keV poor clusters. Green stars are for $kT<$2 keV groups.
Crosses represent upper limits in both axes.
The lower panel shows the histogram for all BCGs (upper limits included)
with two classes marked, while the histogram in red is for $L_{\rm 1.4GHz} > 10^{24}$ W Hz$^{-1}$
BCGs. This histogram can be regarded as a raw {\bf cool-core distribution function}.
There are two classes of BCG cool cores (shown in
orange ellipses): the large-cool-core (LCC) class and the corona class.
Above $L_{\rm 1.4GHz}$ of 2$\times10^{23}$ W Hz$^{-1}$, every BCG has a confirmed cool core,
either in the LCC class or in the corona class (see \citep{sun09} for detail).
Radio AGNs in the corona class can be as luminous as those in LCC class.
{\bf Right}: The electron density of ESO~137-006's corona at the Bondi radius
(62 pc) vs. the slope of the density profile in the innermost bin ($n_{\rm e} \propto r^{-\alpha}$).
The solid line is the relation constrained from the observed normalization of the innermost bin.
The dashed line is for the relation when the normalization
is doubled, which should over-estimate the PSF correction.
The dotted line is from the boundary condition of the density in the second innermost bin
(see Fig. 7 of \citep{sun09}). The intersections of these lines imply the density at the Bondi
radius as $\sim$ 1.5 cm$^{-3}$.
}
\end{figure}

\section{Implications}

While a detailed discussion is presented in \citep{sun09}, we summarize some
implications here:
1) Small coronae, easily overlooked or misidentified as X-ray AGN at $z>$0.1, are
mini-cool-cores in groups and clusters.
They can trigger strong radio outbursts long before large
cool cores are formed. The triggered outbursts can destroy embryonic large
cool cores and thus provide another mechanism besides mergers to prevent formation of
large cool cores. The outbursts triggered by coronae can
also inject extra entropy into the ICM and modify the ICM properties in systems
without large cool cores. 
2) Luminous coronae may be able to power their radio AGN through Bondi accretion,
while the hot accretion may not work for faint coronae in less massive galaxies.
However, a complete inventory of cold gas in embedded
coronae is required to address the question of the accretion mode.
3) While coronae may trigger radio AGN, strong outbursts have to deposit
little energy inside coronae to keep them intact.
Thus, it is unclear whether coronae are decoupled from the radio feedback
cycle.
4) The existence of coronae around strong radio AGN in groups and clusters
also affects the properties of radio jets, e.g., extra pressure to decelerate jets
and maintain the collimation of jets. Its small size also allows the bulk of the
jet energy to transfer to the outskirts of the system, while radio sources
can be strangled in large cool cores.

We want to further stress the importance of small X-ray cool cores like coronae.
There has been a lot of attention to understand large cool cores like Perseus's.
However, most of local massive galaxies (e.g., more luminous than $L_{*}$) are
not in large cool cores. Small X-ray cool cores like coronae are more typical
gaseous atmosphere that actually matters in the evolution of massive galaxies to
explain their colors and luminosity function.
There has also been a lot of discussions on ``bimodality'' of the cluster/group gas cores
(e.g., \citep{cav09}), basically cool cores (with low entropy and power-law
distribution) and noncool cores (with high entropy and flat distribution).
However, {\em one should not misunderstand the bimodality of the ICM entropy at $\geq$ 5 - 10 kpc
scales with the bimodality of the gas entropy at $\sim$ Bondi radius of the central
SMBH.}
The formal ``bimodality'' has been confirmed (e.g., \citep{cav09}), while
we now know that the properties of the gaseous atmosphere around the central SMBH are not necessarily
correlated with the gas properties at larger scales (e.g., $r \geq$ 5 - 10 kpc).
A related basic question is {\em how to define an X-ray cool core}. If we
only care about the evolution of the ICM core around the BCG, these small ISM cool
cores can be ignored. However, they really matter if we care about the radio
AGNs that are important for structure formation.
This is why we prefer a cool-core distribution function as shown in Fig. 2, although
the low-$L_{\rm X}$ end is not easy to be constrained.

\section{Warning on the derivation of Bondi accretion rate}

There have been some attempts to constrain the Bondi accretion rate for SMBHs embedded
in hot gas\citep{allen06}\citep{Bal08}. As the Bondi radius is resolved by \chandra\ in only
$\sim$ 2 galaxies, \citep{allen06} and \citep{Bal08} used the density profile at larger
radii to extrapolate the density at the Bondi radius, assuming $n_{\rm e} \propto r^{-\alpha}$.
Often a very high density was derived (e.g., up to 40 cm$^{-3}$ in \citep{Bal08}),
which makes the required mass-energy conversion efficiency
as low as 1\% \citep{Bal08}. This approach is too simplified. The power-law fits by \citep{Bal08}
produce $\alpha > 1.5$ in many cases, which implies an infinite integrated luminosity within the
central 1 kpc or so. In fact, these previous work did not compare the emission predicted
by their density model with the observed flux. For example, \citep{Bal08} derived
$n_{\rm e}$=42 cm$^{-3} (r/ 32 {\rm pc})^{-1.7}$ for 3C~296. Even assuming the density not increasing
anymore inside the Bondi radius (32 pc), the predicted luminosity within 1 kpc from the above
density model is still 13 times the observed value.
A better approach was used in \citep{sun09} for ESO~137-006,
also shown in Fig. 2. Our approach also assumes a power-law density profile
inside the innermost bin (but other models can also be used).
However, the index is constrained from the observed flux of the
innermost bin and a boundary condition of the density derived in the second innermost bin,
instead of a simple extrapolation from large radii.
Of course the PSF correction is required, which is only approximated in Fig. 2.
Nevertheless, this method effectively constrains the density at the Bondi radius
for ESO~137-006, $\sim$ 1.5 cm$^{-3}$, while an extrapolation from the density profile
at 0.6 - 1.6 kpc would predict a density of 9.1 cm$^{-3}$ at the Bondi radius.
Clearly, many Bondi accretion rates estimated by \citep{Bal08} are largely over-estimated.



\bibliographystyle{aipproc}   

\newcommand\apj{ApJ}
\newcommand\apjs{ApJS}
\newcommand\araa{ARAA}
\newcommand\apjl{ApJ}
\newcommand\mnras{MNRAS}
\newcommand\aap{A\&A}
\newcommand\nat{Nature}

\bibliography{msun}


\end{document}